\documentclass[prl,twocolumn,superscriptaddress]{revtex4}
\usepackage{epsfig}
\usepackage{times}
\usepackage{amsmath}
\usepackage{amsfonts}
\usepackage{amssymb}
\usepackage[usenames]{color}
\usepackage{graphicx}

\newcommand{\beq}{\begin{equation}}
\newcommand{\eeq}{\end{equation}}
\newcommand{\beqa}{\begin{eqnarray}}
\newcommand{\eeqa}{\end{eqnarray}}

\begin{document}
\title{Entanglement generation between unstable optically active
qubits without photodetectors}
\author{Yuichiro Matsuzaki}
\affiliation{Department of Applied Physics/COMP, Aalto University, P.O. Box 14100, FI-00076 AALTO, Finland}
\author{Paolo Solinas
}
\affiliation{Department of Applied Physics/COMP, Aalto University,
P.O. Box 14100, FI-00076 AALTO, Finland}
\author{Mikko M{\"o}tt{\"o}nen
}
\affiliation{Department of Applied Physics/COMP, Aalto University,
P.O. Box 14100, FI-00076 AALTO, Finland}
\affiliation{Low Temperature Laboratory, Aalto University, P.O. Box 13500, FI-00076 AALTO, Finland}

\begin{abstract}

 
We propose a robust deterministic scheme to generate entanglement at high fidelity
 without the need of photodetectors
 even for quantum bits, qubits, with extremely poor optically active states.
 Our protocol employs stimulated Raman adiabatic
passage for population transfer without actually exciting the system.
 Furthermore, it is found to be effective
 even if the environmental decoherence rate is of the same order
 of magnitude as the atom--photon coupling frequency. Our scheme holds potential to solve entanglement generation problems, e.g., in distributed quantum computing.
\end{abstract}

\maketitle
One of the key challenges for practical quantum computing is
scalability.
Recently, an approach referred to as
 \emph{distributed quantum information processing} has been suggested to solve this problem
 \cite{CEHM01a,BK01a,BPH01a,BKPV01a,FZLGX01a,YSJ01a}.
 In this scheme,
 a scalable computer is constructed
  from a network of small devices, each composing of a single or a few quantum bits, qubits.
 Importantly, the relatively long distance
 between the qubits renders it feasible to
 address the individual qubits and to suppress decoherence caused by unknown
 qubit--qubit interactions.
To construct the network,
inter-node entanglement is necessary, and
many proposals of entanglement generation
using a photon
to mediate interactions
between the qubits have been suggested
  \cite{CEHM01a,BK01a,BPH01a,BKPV01a,PRAmatsuzaki2010distributed}.
  Due to weak interactions with its environment,  a photon seems an
  ideal candidate for a flying qubit to generate such shared
  entanglement between computation nodes.

However, such remote entanglement can be degraded by error sources: imperfect photodetection and unstable optically
excited states.
 Although there are many proposals to perform
 entanglement generation,
 in most cases, the use of photodetectors for measuring the emitted photons
 is inevitable,
 and imperfections of detectors causes a
 significant loss of the fidelity.
The first
experimental realization to perform entanglement generation between macroscopically distant
atoms has been reported in Ref.~~\cite{MMOYMDM1a}. The fidelity in this probabilistic method is only $0.63$ after post selection, limited mainly by the dark counts of the photodetectors.
In general, it is usually necessary to involve optical
transitions into an excited state of the qubit for the entanglement generation. Such an excited state is
prone to decoherence due to the strong coupling with the environment \cite{kaerlodahl01a,gerardotetal2008optical}.

A protocol, in which the fidelity of the entanglement generation does not depend on
the imperfections of the photodetectors, has been presented in Ref.~\cite{PRAmatsuzaki2010distributed}. However,
this scheme requires optical transitions and hence becomes sensitive to
the decoherence of the optically excited
state. Although other protocols have been suggested recently in order to
overcome this type of decoherence~\cite{matsuzaki2010entangling,nazir2009overcoming}, the
protocols still require photodetectors to generate
the remote entanglement. Consequently, they are
vulnerable to the imperfect photon detections.
In this Letter, we introduce a
 protocol,
 in which the eventual fidelity is not hindered by either of these
 effects. Surprisingly, without the use of photodetectors, this protocol still
 achieves a high fidelity entanglement even when the environmental
 decoherence rate at the
 excited state is as large as the atom--photon coupling frequency.

 The basic idea of our scheme is to utilize the concept of \emph{single--particle entanglement} suggested by van~Enk \cite{SJvan01a}.
 Suppose that one uses a half-plated mirror to split a single photon into
 two paths, and on each path there is an atom
 prepared in its ground state.
 The photon on each path is focused
 onto the atom to be absorbed. For simplicity, let us
 make the  assumption that the absorption probability is unity.
 Although one of the atoms will be
 excited, it is not possible to know which atom is excited and
 therefore
 a Bell state $|\Psi _e^{(+)}\rangle
 =\frac{1}{\sqrt{2}}(|G\rangle _L|E\rangle _R
 +|E\rangle _L|G\rangle _R)$ can be generated \cite{SJvan01a} where $|G\rangle _i$
 and $|E\rangle _i$ $(i=L,R)$ denote the
 ground state and the excited state of
 the $i$ th atom $(i=L,R)$, respectively.

\begin{figure}[h]
   \begin{center}
       \includegraphics[width=4.0cm]{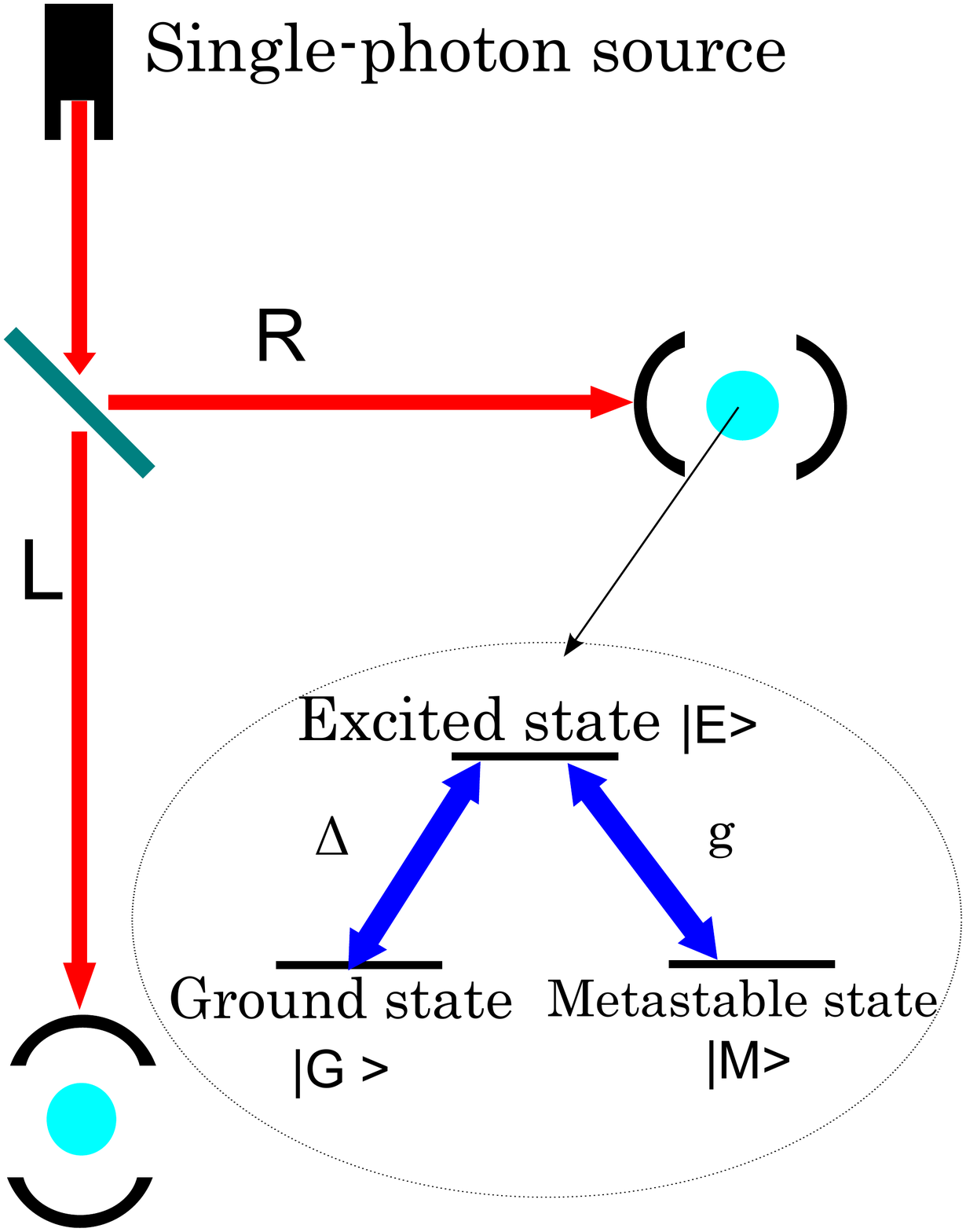}
   \end{center}
  \caption{
Schematic picture of an apparatus for the entanglement generation.
     A half-plated mirror splits a single photon into two paths, and at the end
     of each
     path a qubit is confined in a cavity. Each atom has a
     $\lambda $-type energy level structure, i.e., the
     ground state $|G\rangle $, metastable state $|M\rangle $, and
     excited state $|E\rangle $.  The state $|M\rangle $ is coupled to the excited state $|E\rangle $
 with the cavity coupling strength $g$, and the state $|G\rangle $ is
     coupled to the excited
 state $|E\rangle $ by a classical driving field $\Delta $.
    } \label{schematic}
   \end{figure}

  However, there are several potential sources of errors to
 challenge this simple scheme.
  Firstly, the excited state is usually prone to decoherence
 \cite{NKJ01a,gerardotetal2008optical}.
 Secondly, a low absorption probability is a serious source of error
 in this simple protocol, since
 the interaction between the photon and atom in a free space is weak.
 To overcome these difficulties, our scheme involves a qubit in a cavity forming a lambda-system with a ground state $|G\rangle
 $, an exited state $|E\rangle $, and a metastable state $|M\rangle $ as
 shown in Fig.~\ref{schematic}.
 The state $|M\rangle $ is coupled to the excited state $|E\rangle $
 through the cavity coupling strength $g$ and we induce Rabi
 oscillations between the states $|G\rangle $ and $|E\rangle $ by a laser.
 Initially, the state is prepared to be $|M\rangle _L|M\rangle _R
 |\text{vac}\rangle $ where $|\text{vac}\rangle $ denotes a vacuum state for the cavity photons.
 A single photon split by a half mirror is focused onto the two cavities
 which lie symmetrically.
 As a result, we have $|M\rangle _L|M\rangle _R
 \frac{1}{\sqrt{2}}(\hat{a}^{\dagger }_L+\hat{a}^{\dagger }_R)|\text{vac}\rangle $,
 where $\hat{a}^{\dagger }_i$ $(i=L,R)$ denotes a
 creation operator of a cavity mode
 at the $i$th atom.
 Importantly, by ramping adiabatically off the classical field at each cavity, the
 population of the state $|M\rangle \hat{a}^{\dagger }|\text{vac}\rangle$ can be transferred to the state
 $|G\rangle |\text{vac}\rangle $ essentially without populating
 the excited state
 while the state $|M\rangle|\text{vac}\rangle$ remains intact~\cite{2scully1997quantumbook}. 
 Thus the state evolves into a state $\frac{1}{\sqrt{2}}(|G\rangle _L|M\rangle _R
  +|M\rangle _L|G\rangle _R)|\text{vac}\rangle $. Similar adiabatic transfer between a photon excitation and a single atom in a cavity has been demonstrated experimentally in Ref.~\cite{boozer2007reversible}.
 We stress that, in our scheme, neither a use of photodetectors nor an optical
 transition to the excited state is necessary, and therefore this protocol is robust against
 typical errors caused by the imperfections.

 In the above description, we assumed that the ramp of the classical driving field is
 so slow that
 an optical transition to the excited state does not occur.
Furthermore, the photon loss during this process
 is negligible with a high-Q cavity.
 However, there is a non-zero
 probability to excite the state for a finite operation time, which reduces the fidelity of the entanglement generation.
 In order to include this effect, it is necessary to present a more
 detailed analysis.

 Let us consider first the case of a single-qubit system to study the
 effect of the decoherence of the excited state and the non-adiabaticity. Below, we generalize to the actual two-qubit case for entanglement generation.
 The Hamiltonian of the system is given by
 \begin{eqnarray}
  H_{\text{sys}}(t)&=&H_S+H_I,
   \nonumber \\
  H_S&=&\epsilon |e\rangle \langle e|
   +\omega_c\hat{a}^{\dagger}\hat{a},
   \nonumber \\
  H_I&=&g(|1\rangle \langle e|+|e\rangle
   \langle 1|)
   +\Delta_t
    \cos (\omega t)\Big{(}|e\rangle \langle 0|+|0\rangle
   \langle e|\Big{)},\ \ \ \ \nonumber
 \end{eqnarray}
   where we have defined $|0\rangle =|G\rangle |\text{vac}\rangle $, $|1\rangle = |M\rangle \hat{a}^{\dagger
   }|\text{vac}\rangle $, $|e\rangle =|E\rangle |\text{vac}\rangle $, and set $\hbar =1$ for simplicity.
 Here,  $\Delta _t$ is the amplitude of the time dependent driving field, $\omega$ is its frequency, $\omega
 _c$ is the frequency of the
 cavity mode, $\epsilon $ denotes the energy of the atomic transition,
 and $g$ is the coupling constant of the standard
 Jaynes-Cummings model.
 For simplicity,
 we assume that the driving field, cavity mode, and atomic
 transition are resonant so that $\omega =\omega_c=\epsilon$.
 Since the classical field is ramped adiabatically off, the time
 dependent strength of the field can be represented as $\Delta _t=\Delta
 \cos ^2 (at)$  for $0\leq t\leq \frac{\pi }{2a }$
 where $\frac{1}{a}$ denotes a time scale of the variation of the Hamiltonian.
 In a rotating frame with the laser frequency, which is characterized by a unitary
 operation $e^{i\omega t |0\rangle \langle 0|}$ \cite{kok2010introduction}, we obtain
  \begin{eqnarray}
  H_S&\simeq &\epsilon (|0\rangle \langle 0|+|e\rangle \langle
   e|+|1\rangle \langle 1|),\nonumber \\
  H_I&\simeq &g(|1\rangle \langle e|+|e\rangle
   \langle 1|)
   +\frac{\Delta \cos ^2(at)}{2}(|e\rangle \langle 0|+|0\rangle
   \langle e|),\ \ \ \ \nonumber
 \end{eqnarray}
 where we have employed the rotating wave
 approximation.
 A convenient basis to describe the dynamics of this system is the
 adiabatic basis, defined as $H_{\text{sys}}(t)|E_{\tilde n}(t)\rangle =E_{\tilde n}(t)|E_{\tilde n}(t)\rangle $.
   Also, in an appropriate frame to rescale the energy,
  we can assume
  $\epsilon =0$
  without loss of generality.

Since we have only one excitation in the system,
the eigenstates become
    \begin{eqnarray}
  |E_{\tilde{0}}\rangle
  &=\frac{1}{\sqrt{\Delta ^2\cos ^4 (at)+4g^2 }}(-2g|0\rangle +\Delta
   \cos ^2(at)|1\rangle
   )\label{eq01} \\
  |E_{\tilde{e}}\rangle
  &=\frac{\frac{1}{\sqrt{2}}\Delta
   \cos ^2(at) }{\sqrt{4g^2+\Delta ^2 \cos
  ^4(at)}} (|0\rangle +\frac{2g}{\Delta
   \cos ^2(at)}|1\rangle )
   -\frac{1}{\sqrt{2}}|e\rangle 
   \label{eq02} \\
   |E_{\tilde{1}}\rangle
  &=\frac{\frac{1}{\sqrt{2}}\Delta
   \cos ^2(at)
   }{\sqrt{4g^2+\Delta ^2 \cos
  ^4(at)}} ( |0\rangle +\frac{2g}{\Delta
   \cos ^2(at)}|1\rangle )
   +\frac{1}{\sqrt{2}}|e\rangle. \label{eq03}
 \end{eqnarray}
 The eigenvalues are $E_{\tilde{0}}(t)=0 $,
 $E_{\tilde{e}}(t)=
-\frac{\sqrt{4g^2+\Delta ^2 \cos ^4(at)}}{2}$, and
$E_{\tilde{1}}(t)=\frac{\sqrt{4g^2+\Delta ^2 \cos ^4(at)}}{2}$.
To be able to study the joint effect of decoherence and the adiabatic evolution, we aim to map the dynamical system into a time-independent one~\cite{pekola2010decoherence,solinasetal2010adiabatically}. To this end, we define a transformed system density operator $\tilde\rho_\textrm{sys}=\hat{D}^\dagger\rho_\textrm{sys}\hat{D}$, the time evolution of which is governed by the effective Hamiltonian $H_{\text{eff}}=\tilde{H}(t)+w$. Here the Hamiltonian
 $\tilde{H}(t)=\hat{D}^{\dagger}H(t)\hat{D}=\sum_{k=0,e,1} E_{\tilde{k}}|k\rangle \langle k|$ is diagonal in the time-independent basis $\{|k\rangle\}$, the operator $\hat{D}=\sum_{k=0,e,1} |E_{\tilde{k}}\rangle \langle k|$, and $w=-i
 \hat{D}^{\dagger}\frac{d\hat{D}}{dt}=-\frac{2\sqrt{2}ig\cdot a \cdot \Delta \cos (at)\sin (at)}{4g^2+\Delta ^2 \cdot \cos
 ^4(at)} (|0\rangle  \langle 1|+ |0\rangle
 \langle e|-|1\rangle \langle 0|- |e\rangle \langle 0|)$.
 The remaining time dependence is manifested in the correction term $w$ which tends to excite the system.
 There are a few possible strategies to treat this correction term.
 The simplest scheme is to disregard the correction term completely, which can be valid
 only in the adiabatic limit. To include the
 non-adiabatic correction to the lowest order, it is possible to perform another transformation corresponding to the
 super-adiabatic basis~\cite{salmilehtoetal2010adiabatically} as we will do in
 this paper.
 Namely, we diagonalize the Hamiltonian $\tilde{H}(t) +w$ using the
 first-order
 perturbation theory on the correction term~\cite{salmilehtoetal2010adiabatically}.
 Note that there is no energy shift in this order of the
perturbation theory.
The approximate eigenstates of the effective Hamiltonian in the untransformed system, i.e, eigenstates of $\hat{D}(\tilde{H}(t)+w)\hat{D}^\dagger$, 
 are expressed with the help of Eqs.~\eqref{eq01}--\eqref{eq03} as
  \begin{eqnarray}\label{eq1}
  |E_{\tilde{\tilde{0}}}\rangle = -z_t|0\rangle +y_t|1\rangle
   -\sqrt{2}x_t|e\rangle \ \ \ \ \ \ \ \ \ \ \ \ \ \ \ \ \ \
   \ \ \ \ \ \ \ \ \ \ \ \ \ \ \ \ \ \
    \\ \label{eq2}
 |E_{\tilde{\tilde{e}}}\rangle =(\frac{1}{\sqrt{2}}y_t-x_tz_t)|0\rangle
   +(\frac{1}{\sqrt{2}}z_t+x_ty_t)|1\rangle
   -\frac{1}{\sqrt{2}}|e\rangle
   \\ \label{eq3}
 |E_{\tilde{\tilde{1}}}\rangle =(\frac{1}{\sqrt{2}}y_t+x_tz_t)|0\rangle
   +(\frac{1}{\sqrt{2}}z_t-x_ty_t)|1\rangle
   +\frac{1}{\sqrt{2}}|e\rangle   
 \end{eqnarray}
where $x_t=\frac{4\sqrt{2}ig a \Delta \cos (at)\sin (at)}{[4g^2+\Delta ^2 \cos ^4(at)]^{\frac{3}{2}}}$,
$y_t=\frac{\Delta \cos ^2(at)}{\sqrt{4g^2+\Delta ^2 \cos ^4(at)}}$
and $z_t=\frac{2g}{\sqrt{4g^2+\Delta ^2 \cos ^4(at)}}$.
As long as the adiabatic condition
$\frac{|E_{\tilde{0}}-E_{\tilde{1}}|}{a}, \frac{|E_{\tilde{0}}-E_{\tilde{e}}|}{a}\ll
1$
is
satisfied, an initial state $|E_{\tilde{\tilde{0}}}(0)\rangle $ remains
in the state $|E_{\tilde{\tilde{0}}}(t)\rangle $ during the adiabatic
process \cite{messiahquantum}.
Since we have $|E_{\tilde{\tilde{0}}}(0)\rangle \approx |1\rangle $ for
$\frac{g}{\Delta }\ll 1$ and $|E_{\tilde{\tilde{0}}}(\frac{\pi }{2a })\rangle
=-|0\rangle $,
the population transfers from the state
$|1\rangle $ to the state $|0\rangle $ when the driving field is ramped off adiabatically.
In Eq.~\eqref{eq1}, $\sqrt{2}x_t$ denotes the amplitude of the excited
state yielding $P_e= 2|x_t|^2\simeq \frac{64 g^2 a^2\Delta ^2\cos ^2(at)\sin ^2(at)}{[4g^2 +\Delta ^2
     \cos ^4(at)]^3}$ for the excited-state probability. Since we assume that $g\ll\Delta$, the maximum probability is obtained at a point where $\sin(at)\approx 1$. With this approximation, the maximum excited-state probability is given by
     $\frac{25\sqrt{5}}{108}\frac{a^2\Delta }{g^3}$ for $t=\frac{1}{a}\arccos
     \sqrt{\frac{2g}{\sqrt{5}\Delta }}$.
We conclude that the slow ramp rate of the driving field prevents the adiabatically evolving state to have projection on the excited state, hence protecting the system from the decoherence of the excited state.

\begin{figure}[h]
   \begin{center}
       \includegraphics[width=8cm]{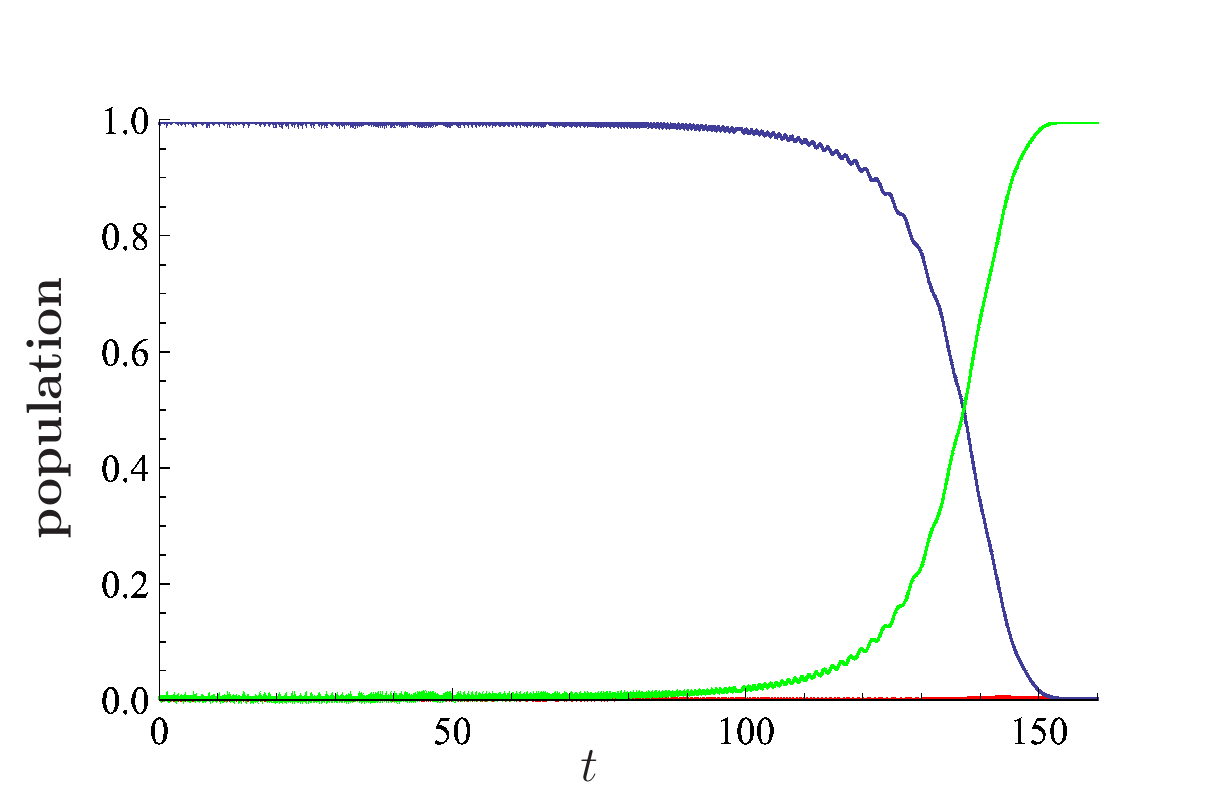}
   \end{center}
  \caption{(Color Online) Population of the state $|0\rangle$ (green ascending curve),
    $|e\rangle $ (red almost constant curve), and $|1\rangle $
    (blue descending curve) during the adiabatic transfer under the effect of
    decoherence at the excited state $|e\rangle $.
    We have employed parameters $a=0.01\times g$, $\Delta=50\times g$, and $\gamma=0.1\times g$. The results are independent of the value of $g$. The population of state $|1\rangle $ is transferred to the target
    state $|0\rangle $ with fidelity
    $0.992$ when the external field is ramped off.
    } \label{population-plot}
   \end{figure}

However, we did not take the decoherence processes explicitly into account above. Especially, the optically excited state
$|e\rangle $ is usually coupled strongly with the environment causing
decoherence of the quantum states \cite{kaerlodahl01a,gerardotetal2008optical}.
Although the population of the excited state should be small in our
scheme because of the slow variation,
we study how the noise degrades the coherence. Therefore, we need to derive a
Markovian master equation.
The total Hamiltonian of the system and the environment is represented
as $ H=H_\text{sys}(t)+H_{\text{int}}+H_{\text{env}}$ where
$H_{\text{env}}$ denotes a bath operator and
$H_{\text{int}}$ denotes the coupling between the
system and the environment.
Since we consider the noise at the excited state, the
coupling between the system and the environment can be represented as
$H_{\text{int}}=\lambda |e\rangle \langle e|\otimes \hat{X}$ where $\hat{X}$ is
the environmental operator.
Equations~\eqref{eq1}--\eqref{eq3} provide us conveniently the transformation operator corresponding to the superadiabatic basis $\hat{D}_s=\sum_{k=0,e,1}|E_{\tilde{\tilde k}}\rangle\langle k|$ which yields the transformed density operator $\tilde{\tilde \rho}_\textrm{sys}=\hat{D}_s^\dagger\rho_\textrm{sys}\hat{D}_s$. In this approximation, the time derivative of $\hat{D}_s$ is neglected and the effective total Hamiltonian becomes $\tilde{\tilde{H}}=\tilde{\tilde{H}}_{\text{sys}}+\tilde{\tilde{H}}_{\text{int}}+H_{\text{env}}$,
where 
$\tilde{\tilde{H}}_{\text{sys}}= \sum_{j=0,1,e}E_{\tilde{j}}|j\rangle \langle j|$, 
$\tilde{\tilde{H}}_{\text{int}}
=|\tilde{\tilde \phi} \rangle \langle \tilde{\tilde \phi}| \otimes \hat{X}$, and $|\tilde{\tilde \phi}\rangle=\sqrt{2}x _t|0\rangle-\frac{1}{\sqrt{2}}|e\rangle+\frac{1}{\sqrt{2}}|1\rangle +O(|x_t|^2)$.
Under the assumption of a white noise spectrum of the environment,
we integrate the von Neumann equation for the total density matrix, trace out the environment, and arrive at the master equation
 \begin{eqnarray}
  \frac{d\tilde{\tilde{\rho
   }}_{\text{sys}}}{dt}=-i[\tilde{\tilde{H}}_{\text{sys}},\tilde{\tilde{\rho
   }}_{\text{sys}} (t)]-\frac{\gamma}{2}[\hat{L},[\hat{L},\tilde{\tilde{\rho }}_{\text{sys}}]],
 \end{eqnarray}
where $\gamma $ denotes the decoherence rate and
$\hat{L}$ denotes a Lindblad operator defined as
$\hat{L}=|\tilde{\tilde \phi} \rangle \langle \tilde{\tilde \phi}|$.
Note that this noise operator can cause unwanted transitions from the adiabatically evolving state $|E_{\tilde{\tilde 0}}\rangle\langle E_{\tilde{\tilde 0}}|\approx \rho_\textrm{sys}$ to the other
states.
  We solve this master equation numerically and show the population
  of each state $P_{j}(t)=\langle j|\rho_\textrm{sys} (t)|j\rangle $ ($j=0,1,e$) in
  Fig. \ref{population-plot}.
  Starting from the state $\rho_\textrm{sys}(0)=|1\rangle\langle 1|$,
  we have achieved $P_0(\frac{\pi }{2a})=0.992$ for $a=0.01\times g$,
  $\Delta =50\times g$, and  $\gamma =0.1\times g$, which shows almost perfect transfer
  from the initial state $|1\rangle\langle 1| $ to the target state $|0\rangle\langle 0| $
  even under the effect of decoherence at the excited state.

  To estimate the effect of dephasing of the adiabatically evolving state due to the decoherence at the excited
  state, we consider an untransformed initial state
  $|\psi _0\rangle =\frac{1}{\sqrt{2}}|E_{\tilde{\tilde{0}}}\rangle
  +\frac{1}{\sqrt{2}}|r\rangle $ where $|r\rangle $ denotes a reference
  state which is not coupled with any other states.
  Here, we assume that, due to the strong driving field $\frac{g}{\Delta
  }\ll 1$, the effect of an imperfect initialization is negligible.
  Since we are interested in an early time stage of the decoherence process, we approximate
  \begin{eqnarray}
   \tilde{\tilde{\rho }}_{\text{sys},I}(t)\approx \hat {D}_s^\dagger|\psi _0\rangle \langle \psi _0
    |\hat{D}_s+\frac{\gamma
    }{2}\int_{0}^{t}dt'[\hat{L}_I(t'),[\hat{L}_I(t'),\tilde{\tilde{\rho
    }}_{\text{sys},I}(0)]], \nonumber
  \end{eqnarray}
  where we substitute $\tilde{\tilde{\rho }}_{\text{sys},I}(t)$ with
  $\tilde{\tilde{\rho }}_{\text{sys},I}(0)$ in the integrand. We have
  employed the interaction picture defined for the operators as
$ \hat{O}_I(t)=\hat{U}^{\dagger
  }_S\hat{O}(t)U_{S}(t)$, where
  $U_S(t)=e^{-i\int_{0}^{t}dt'\tilde{\tilde{H}}_{\text{sys}}(t')}$ is the system
  time-evolution operator.

  Thus, the fidelity can be calculated as follows:
  \begin{eqnarray}
   F&=&{}_S\langle \psi (t)|\tilde{\tilde{\rho }}_{\text{sys},S}(t)|\psi (t)\rangle _S \nonumber \\
   &\approx &1-\gamma \int_{0}^{t}|x_t|^2 +O(|x_t|^3)
  \end{eqnarray}
   where
   $|\psi (t)_{\text{}}\rangle _{S}$ denotes the transformed adiabatically evolving
   state at a time $t$ for vanishing decoherence.
  Since we have $\frac{g}{\Delta }\ll 1$,  $|x_t|$ is negligible
except for a
time region satisfying $0<\Delta \cos ^2(at) < cg$, where $c$ is some constant. In this time region, we
  have  $0<\frac{\pi
  }{2a}-t < \frac{n}{a}\sqrt{\frac{g}{\Delta }}$, where $n$ is a constant that depends only on $c$, not on $a$. Hence we obtain
  $1-F  \approx  \gamma \int_{\frac{\pi}{2a}-\frac{n}{a}\sqrt{\frac{g}{\Delta
}}}^{\frac{\pi}{2a}}|x_{t'}|^2dt' = \frac{1}{2}\gamma \int_{\frac{\pi}{2a}-\frac{n}{a}\sqrt{\frac{g}{\Delta
}}}^{\frac{\pi}{2a}}P_edt'
\leq \frac{25\sqrt{5}}{216} \gamma
\frac{n}{a}\sqrt{\frac{g}{\Delta }} \frac{a^2\Delta }{g^3}$.
Therefore we obtain $1-F=O(a)$.
As expected, the harmful effect of the decoherence decreases with $a$.

\begin{figure}[h]
   \begin{center}
       \includegraphics[width=8cm]{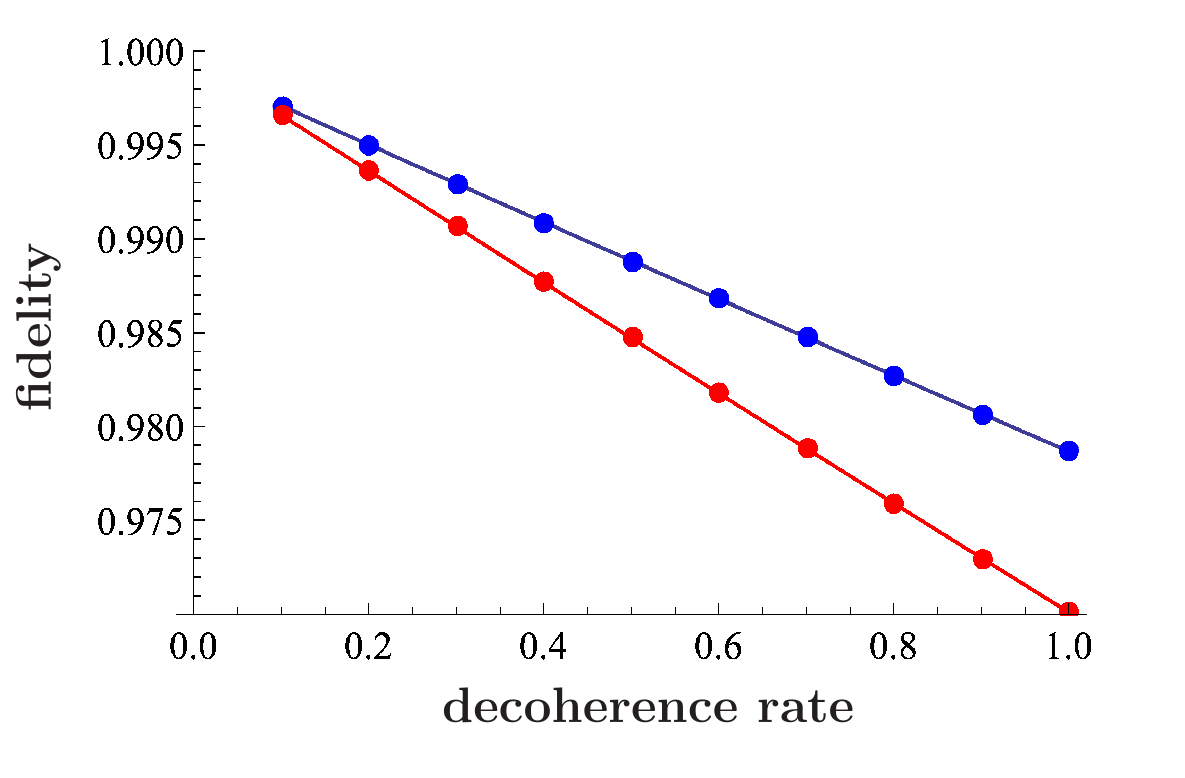}
   \end{center}
  \caption{Fidelity of the proposed entanglement generation scheme as a function of
      the decoherence rate of the environment, $\gamma$, with (top curve) and without (bottom curve)
      rotating wave approximation.
      We have fixed the parameters as $a=0.01\times g$, and $\Delta =70\times g$.
    } \label{scale-fidelity}
   \end{figure}

 Let us generalize the above results
 to the two-qubit case. Since we consider independent decoherence processes for the qubits, the
 master equation for the two-qubit system is $  \frac{d\tilde{\tilde{\rho
   }}_{\text{sys}}}{dt}=\sum_{j=L,R}-i[\tilde{\tilde{H}}^{(j)}_{\text{sys}},\tilde{\tilde{\rho
   }}_{\text{sys}}
 ]-\frac{\gamma}{2}[\hat{L}_{j},[\hat{L}_{j},\tilde{\tilde{\rho
 }}_{\text{sys}}]]$.
We solve the
 master equation numerically using the initial state $|M\rangle _{L}|M\rangle _{R}\frac{1}{\sqrt{2}}(\hat{a}^{\dagger
 }_L+\hat{a}^{\dagger }_R)|\textrm{vac}\rangle$, and show the fidelity $\langle \psi
 _{\text{Bell}}|\rho_\textrm{sys} (\frac{\pi }{2a})|\psi _{\text{Bell}} \rangle $ in Fig.~\ref{scale-fidelity}.
 Here, $|\psi _{\text{Bell}}
 \rangle =\frac{1}{\sqrt{2}}(|M\rangle _{L}|G\rangle _R+
|G\rangle _{L}|M\rangle _R)|\text{vac}\rangle $.
For reference, we show also the fidelity with neither rotating wave approximation nor transformation corresponding to the superadiabatic basis. The two methods give similar results which verifies that our entanglement generation takes place at high fidelity independent of the theoretical approach.
 Surprisingly, even when the environmental decoherence rate $\gamma $ is
 as large as the atom--photon coupling frequency $g$, the protocol still
 generates high-fidelity entanglement.

 Photon loss is one of the major problems in the most of previous
 entanglement generation protocols
 \cite{CEHM01a,LBK01a,BPH01a,BKPV01a,Fengetal01a,laddetal2006hybrid}.
In our case, it can also degrade the fidelity since the photon might be lost before coming
 in the cavity, while in or entering the cavity, or the pulse
generated by the single-photon source may fail to provide a photon~\cite{LM01a}. The state with photon loss can be described by
 $\rho _{LR}=(1-P_{\text{loss}})|\psi_{\text{Bell}} \rangle \langle \psi_{\text{Bell}}
 |+P_{\text{loss}}|M\rangle_{L}|M\rangle_{R}\,_L\langle M|\,_M\langle M|$ even for a perfect
 adiabatic transfer
 where $P_{\text{loss}}$
 is a probability to lose the photon.
     Fortunately, this is of the same form as the resource state considered
 in Refs \cite{CB01a,PRAmatsuzaki2010distributed, matsuzaki2011mobile}.
If we have an ancillary qubit near the actual qubit at each location, we can perform an
efficient two round distillation protocol as follows:
 It is possible to utilize
   the state $\rho _{LR}$  as a resource to perform the parity projection
   on the ancillary qubits.
Although this parity projection is
   imperfect due to the photon loss, by generating the state $\rho
   _{LR}$ again through another adiabatic transfer, one can utilize the second state for
   performing the second parity projection on the ancillary qubits in
   order to obtain a perfect Bell state. The measurement results
   of the two parity projections let us know whether entanglement between
   the ancillary qubits is generated
   or not, and
the
success probability is calculated $P_s=\frac{(1-P_{\text{loss}})^2}{2}$~\cite{CB01a}.
Note that the parity projection is one of the most commonly proposed
  entanglement generation methods~\cite{BK01a, CB01a,PRAmatsuzaki2010distributed}
  to make a two dimensional cluster state
   for quantum
   computation~\cite{Raussendorf:2001p368}.
   Moreover, the joint effect of the photon loss and decoherence in this
distillation protocol has been studied~\cite{matsuzaki2011mobile}, and
has been shown to be negligible
for weak decoherence. These results show that this distillation
protocol makes our scheme robust against
the photon loss by using ancillary qubits near the optically active qubits.
   Such ancillary qubits are available with some species of nanostructures such as the nitrogen--vacancy centers in diamond~\cite{DCJTMJZHL01a,NMRHWYJGJW01a}.

In conclusion, we propose a deterministic scheme to generate entanglement between unstable optically active
qubits.
Our method is designed to function without the
need for photodetectors which are typically major sources of error.
With the help of adiabatic transfer, we
entangle the qubits without exciting the unstable states, which renders our
proposal extremely robust against decoherence.
The authors thank T.\ Close and B.\ Lovett
for useful discussions. We acknowledge Academy of Finland, Emil
Aaltonen Foundation, and Centre for International Mobility for financial support.




\begin{thebibliography}{29}
\expandafter\ifx\csname natexlab\endcsname\relax\def\natexlab#1{#1}\fi
\expandafter\ifx\csname bibnamefont\endcsname\relax
  \def\bibnamefont#1{#1}\fi
\expandafter\ifx\csname bibfnamefont\endcsname\relax
  \def\bibfnamefont#1{#1}\fi
\expandafter\ifx\csname citenamefont\endcsname\relax
  \def\citenamefont#1{#1}\fi
\expandafter\ifx\csname url\endcsname\relax
  \def\url#1{\texttt{#1}}\fi
\expandafter\ifx\csname urlprefix\endcsname\relax\def\urlprefix{URL }\fi
\providecommand{\bibinfo}[2]{#2}
\providecommand{\eprint}[2][]{\url{#2}}

\bibitem[{\citenamefont{Cirac et~al.}(1999)\citenamefont{Cirac, Ekert, Huelga,
  and Macchiavello}}]{CEHM01a}
\bibinfo{author}{\bibfnamefont{J.~I.} \bibnamefont{Cirac}},
  \bibinfo{author}{\bibfnamefont{A.~K.} \bibnamefont{Ekert}},
  \bibinfo{author}{\bibfnamefont{S.~F.} \bibnamefont{Huelga}},
  \bibnamefont{and}
  \bibinfo{author}{\bibfnamefont{C.}~\bibnamefont{Macchiavello}},
  \bibinfo{journal}{Phys. Rev. A} \textbf{\bibinfo{volume}{59}},
  \bibinfo{pages}{4249} (\bibinfo{year}{1999}).

\bibitem[{\citenamefont{Barrett and Kok}(2005)}]{BK01a}
\bibinfo{author}{\bibfnamefont{S.~D.} \bibnamefont{Barrett}} \bibnamefont{and}
  \bibinfo{author}{\bibfnamefont{P.}~\bibnamefont{Kok}},
  \bibinfo{journal}{Phys. Rev. A} \textbf{\bibinfo{volume}{71}},
  \bibinfo{pages}{060310} (\bibinfo{year}{2005}).

\bibitem[{\citenamefont{Browne et~al.}(2003)\citenamefont{Browne, Plenio, and
  Huelga}}]{BPH01a}
\bibinfo{author}{\bibfnamefont{D.~E.} \bibnamefont{Browne}},
  \bibinfo{author}{\bibfnamefont{M.~B.} \bibnamefont{Plenio}},
  \bibnamefont{and} \bibinfo{author}{\bibfnamefont{S.~F.}
  \bibnamefont{Huelga}}, \bibinfo{journal}{Phys. Rev. Lett}
  \textbf{\bibinfo{volume}{91}}, \bibinfo{pages}{067901}
  (\bibinfo{year}{2003}).

\bibitem[{\citenamefont{Bose et~al.}(1999)\citenamefont{Bose, Knight, Plenio,
  and Vedral}}]{BKPV01a}
\bibinfo{author}{\bibfnamefont{S.}~\bibnamefont{Bose \emph{et al.}}},
  \bibinfo{journal}{Phys. Rev. Lett} \textbf{\bibinfo{volume}{83}},
  \bibinfo{pages}{5158} (\bibinfo{year}{1999}).

\bibitem[{\citenamefont{Feng et~al.}(2003{\natexlab{a}})\citenamefont{Feng,
  Zhang, Li, Gong, and Xu}}]{FZLGX01a}
\bibinfo{author}{\bibfnamefont{X.~L.} \bibnamefont{Feng} \emph{et al.}},
  \bibinfo{journal}{Phys. Rev. Lett} \textbf{\bibinfo{volume}{90}},
  \bibinfo{pages}{217902} (\bibinfo{year}{2003}{\natexlab{a}}).

\bibitem[{\citenamefont{Matsuzaki
  et~al.}(2010{\natexlab{a}})\citenamefont{Matsuzaki, C.Benjamin, and
  Fitzsimons}}]{YSJ01a}
\bibinfo{author}{\bibfnamefont{Y.}~\bibnamefont{Matsuzaki}},
  \bibinfo{author}{\bibfnamefont{S.}~\bibnamefont{C.Benjamin}},
  \bibnamefont{and}
  \bibinfo{author}{\bibfnamefont{J.}~\bibnamefont{Fitzsimons}},
  \bibinfo{journal}{Phys. Rev. Lett} \textbf{\bibinfo{volume}{104}},
  \bibinfo{pages}{4} (\bibinfo{year}{2010}{\natexlab{a}}).

\bibitem[{\citenamefont{Matsuzaki
  et~al.}(2010{\natexlab{b}})\citenamefont{Matsuzaki, Benjamin, and
  Fitzsimons}}]{PRAmatsuzaki2010distributed}
\bibinfo{author}{\bibfnamefont{Y.}~\bibnamefont{Matsuzaki}},
  \bibinfo{author}{\bibfnamefont{S.}~\bibnamefont{Benjamin}}, \bibnamefont{and}
  \bibinfo{author}{\bibfnamefont{J.}~\bibnamefont{Fitzsimons}},
  \bibinfo{journal}{Phys. Rev. A} \textbf{\bibinfo{volume}{82}},
  \bibinfo{pages}{010302} (\bibinfo{year}{2010}{\natexlab{b}}).

\bibitem[{\citenamefont{Moehring et~al.}(2007)\citenamefont{Moehring, Maunz,
  Olmschenk, Younge, Matsukevich, Duan, and Monroe}}]{MMOYMDM1a}
\bibinfo{author}{\bibfnamefont{D.~L.} \bibnamefont{Moehring \emph{et al.}}},
  \bibinfo{journal}{Nature} \textbf{\bibinfo{volume}{449}}, \bibinfo{pages}{68}
  (\bibinfo{year}{2007}).

\bibitem[{\citenamefont{Kaer and {\it{et al}}}(2010)}]{kaerlodahl01a}
\bibinfo{author}{\bibfnamefont{P.}~\bibnamefont{Kaer \emph{et al.}}}, \bibinfo{journal}{Phys. Rev.
  Lett.} \textbf{\bibinfo{volume}{104}}, \bibinfo{pages}{157401}
  (\bibinfo{year}{2010}).

\bibitem[{\citenamefont{Gerardot and {\it{et
  al}}}(2008)}]{gerardotetal2008optical}
\bibinfo{author}{\bibfnamefont{B.}~\bibnamefont{Gerardot \emph{et al.}}}, \bibinfo{journal}{Nature}
  \textbf{\bibinfo{volume}{451}}, \bibinfo{pages}{441} (\bibinfo{year}{2008}).

\bibitem[{\citenamefont{Matsuzaki
  et~al.}(2010{\natexlab{c}})\citenamefont{Matsuzaki, Benjamin, and
  Fitzsimons}}]{matsuzaki2010entangling}
\bibinfo{author}{\bibfnamefont{Y.}~\bibnamefont{Matsuzaki}},
  \bibinfo{author}{\bibfnamefont{S.}~\bibnamefont{Benjamin}}, \bibnamefont{and}
  \bibinfo{author}{\bibfnamefont{J.}~\bibnamefont{Fitzsimons}},
  \bibinfo{journal}{arXiv:1009.4171}
  (\bibinfo{year}{2010}{\natexlab{c}}).

\bibitem[{\citenamefont{Nazir and Barrett}(2009)}]{nazir2009overcoming}
\bibinfo{author}{\bibfnamefont{A.}~\bibnamefont{Nazir}} \bibnamefont{and}
  \bibinfo{author}{\bibfnamefont{S.}~\bibnamefont{Barrett}},
  \bibinfo{journal}{Phys. Rev. A} \textbf{\bibinfo{volume}{79}},
  \bibinfo{pages}{11804} (\bibinfo{year}{2009}).

\bibitem[{\citenamefont{Enk}(2005)}]{SJvan01a}
\bibinfo{author}{\bibfnamefont{S.~J.} \bibnamefont{van Enk}},
  \bibinfo{journal}{Phys. Rev. A} \textbf{\bibinfo{volume}{72}},
  \bibinfo{pages}{064306} (\bibinfo{year}{2005}).

\bibitem[{\citenamefont{Neumann et~al.}(2009)\citenamefont{Neumann, Kolesov,
  Jacques, Beck, Tisler, Batalov, Rogers, Manson, Balasubramanian, Jelezko
  et~al.}}]{NKJ01a}
\bibinfo{author}{\bibfnamefont{P.}~\bibnamefont{Neumann \emph{et al.}}}, 
\bibinfo{journal}{New J. Phys.}
  \textbf{\bibinfo{volume}{11}}, \bibinfo{pages}{013017}
  (\bibinfo{year}{2009}).

\bibitem[{\citenamefont{Scully and Zubairy}(1997)}]{2scully1997quantumbook}
\bibinfo{author}{\bibfnamefont{M.}~\bibnamefont{Scully}} \bibnamefont{and}
  \bibinfo{author}{\bibfnamefont{M.}~\bibnamefont{Zubairy}},
  \emph{\bibinfo{title}{{Quantum Optics}}} (\bibinfo{publisher}{Cambridge},
  \bibinfo{year}{1997}).

\bibitem[{\citenamefont{Boozer et~al.}(2007)\citenamefont{Boozer, Boca, Miller,
  Northup, and Kimble}}]{boozer2007reversible}
\bibinfo{author}{\bibfnamefont{A.}~\bibnamefont{Boozer \emph{et al.}}},
  \bibinfo{journal}{Phys. Rev. Lett.} \textbf{\bibinfo{volume}{98}},
  \bibinfo{pages}{193601} (\bibinfo{year}{2007}).

\bibitem[{\citenamefont{Kok and Lovett}(2010)}]{kok2010introduction}
\bibinfo{author}{\bibfnamefont{P.}~\bibnamefont{Kok}} \bibnamefont{and}
  \bibinfo{author}{\bibfnamefont{B.}~\bibnamefont{Lovett}},
  \emph{\bibinfo{title}{{Introduction to optical quantum information
  processing}}} (\bibinfo{publisher}{Cambridge Univiversity Press}, \bibinfo{year}{2010}).

\bibitem[{\citenamefont{Pekola and {\it{et al}}}(2010)}]{pekola2010decoherence}
\bibinfo{author}{\bibfnamefont{J.}~\bibnamefont{Pekola \emph{et al.}}}, \bibinfo{journal}{Phys. Rev.
  Lett.} \textbf{\bibinfo{volume}{105}}, \bibinfo{pages}{30401}
  (\bibinfo{year}{2010}).

\bibitem[{\citenamefont{Solinas {\it{et
  al.}}}(2010)}]{solinasetal2010adiabatically}
\bibinfo{author}{\bibfnamefont{P.}~\bibnamefont{Solinas \emph{et al.}}}, \bibinfo{journal}{Phys. Rev. B}
  \textbf{\bibinfo{volume}{82}}, \bibinfo{pages}{134517}
  (\bibinfo{year}{2010}).

\bibitem[{\citenamefont{Salmilehto and {\it{et
  al}}}(2010)}]{salmilehtoetal2010adiabatically}
\bibinfo{author}{\bibfnamefont{J.}~\bibnamefont{Salmilehto \emph{et al.}}}, \bibinfo{journal}{Phys. Rev. A}
  \textbf{\bibinfo{volume}{82}}, \bibinfo{pages}{062112}
  (\bibinfo{year}{2010}).

\bibitem[{\citenamefont{Messiah}(1996)}]{messiahquantum}
\bibinfo{author}{\bibfnamefont{A.}~\bibnamefont{Messiah}},
  \bibinfo{journal}{\emph{Quantum mechanics}}, \bibinfo{pages}{Wiley, New York}
  (\bibinfo{year}{1962}).



\bibitem[{\citenamefont{Lim et~al.}(2005)\citenamefont{Lim, Beige, and
  Kwek}}]{LBK01a}
\bibinfo{author}{\bibfnamefont{Y.~L.} \bibnamefont{Lim}},
  \bibinfo{author}{\bibfnamefont{A.}~\bibnamefont{Beige}}, \bibnamefont{and}
  \bibinfo{author}{\bibfnamefont{L.~C.} \bibnamefont{Kwek}},
  \bibinfo{journal}{Phys. Rev. Lett} \textbf{\bibinfo{volume}{95}},
  \bibinfo{pages}{030505} (\bibinfo{year}{2005}).

\bibitem[{\citenamefont{Feng et~al.}(2003{\natexlab{b}})\citenamefont{Feng,
  Zhang, Li, Gong, and Xu}}]{Fengetal01a}
\bibinfo{author}{\bibfnamefont{X.~L.} \bibnamefont{Feng \emph{et al.}}},
  \bibinfo{journal}{Phys. Rev. Lett} \textbf{\bibinfo{volume}{90}},
  \bibinfo{pages}{217902} (\bibinfo{year}{2003}{\natexlab{b}}).

\bibitem[{\citenamefont{Ladd et~al.}(2006)\citenamefont{Ladd, Van~Loock,
  Nemoto, Munro, and Yamamoto}}]{laddetal2006hybrid}
\bibinfo{author}{\bibfnamefont{T.}~\bibnamefont{Ladd \emph{et al}}},
  \bibinfo{journal}{New J. Phys.} \textbf{\bibinfo{volume}{8}},
  \bibinfo{pages}{184} (\bibinfo{year}{2006}).

\bibitem[{\citenamefont{Lounis and Moerner}(2000)}]{LM01a}
\bibinfo{author}{\bibfnamefont{B.}~\bibnamefont{Lounis}} \bibnamefont{and}
  \bibinfo{author}{\bibfnamefont{W.~E.} \bibnamefont{Moerner}},
  \bibinfo{journal}{Nature} \textbf{\bibinfo{volume}{407}},
  \bibinfo{pages}{491} (\bibinfo{year}{2000}).

\bibitem[{\citenamefont{Campbell and Benjamin}(2008)}]{CB01a}
\bibinfo{author}{\bibfnamefont{E.~T.} \bibnamefont{Campbell}} \bibnamefont{and}
  \bibinfo{author}{\bibfnamefont{S.~C.} \bibnamefont{Benjamin}},
  \bibinfo{journal}{Phys. Rev. Lett.} \textbf{\bibinfo{volume}{101}},
  \bibinfo{pages}{130502} (\bibinfo{year}{2008}).

\bibitem[{\citenamefont{Matsuzaki and Jefferson}(2011)}]{matsuzaki2011mobile}
\bibinfo{author}{\bibfnamefont{Y.}~\bibnamefont{Matsuzaki}} \bibnamefont{and}
  \bibinfo{author}{\bibfnamefont{J.~H.} \bibnamefont{Jefferson}},
  \bibinfo{journal}{arXiv:1102.3121}  (\bibinfo{year}{2011}).

\bibitem[{\citenamefont{Raussendorf and Briegel}(2001)}]{Raussendorf:2001p368}
\bibinfo{author}{\bibfnamefont{R.}~\bibnamefont{Raussendorf}} \bibnamefont{and}
  \bibinfo{author}{\bibfnamefont{H.}~\bibnamefont{Briegel}},
  \bibinfo{journal}{Phys. Rev. Lett.} \textbf{\bibinfo{volume}{86}},
  \bibinfo{pages}{5188} (\bibinfo{year}{2001}).

\bibitem[{\citenamefont{Dutt et~al.}(2007)\citenamefont{Dutt, Childress, Jiang,
  Togan, Maze, Jelezko, Zibrov, Hemmer, and Lukin}}]{DCJTMJZHL01a}



\bibinfo{author}{\bibfnamefont{M.~V.~G.} \bibnamefont{Dutt \emph{et al.}}},
  \bibinfo{journal}{Science} \textbf{\bibinfo{volume}{316}},
  \bibinfo{pages}{1312} (\bibinfo{year}{2007}).

\bibitem[{\citenamefont{Neumann et~al.}(2008)\citenamefont{Neumann, Mizuochi,
  Rempp, Hemmer, Watanabe, Yamasaki, Jacques, Gaebel, Jelezko, and
  Wrachtrup}}]{NMRHWYJGJW01a}
\bibinfo{author}{\bibfnamefont{P.}~\bibnamefont{Neumann \emph{et al.}}},
  \bibinfo{journal}{Science} \textbf{\bibinfo{volume}{320}},
  \bibinfo{pages}{1326} (\bibinfo{year}{2008}).

\end{thebibliography}

\end{document}